\documentclass[10pt,twocolumn,english,aps,pra,superscriptaddress,address,showpacs,showacknowledgments]{revtex4-1}
\usepackage[T1]{fontenc}
\usepackage[latin9]{inputenc}
\usepackage[a4paper]{geometry}
\usepackage[active]{srcltx}
\usepackage{xcolor}
\usepackage{amsmath}
\usepackage{amssymb}
\usepackage{stackrel}
\usepackage{graphicx}
\PassOptionsToPackage{normalem}{ulem}
\usepackage{ulem}



\usepackage{babel}
\begin{document}

\title{Robust Rydberg gate via Landau-Zener control of F\"{o}rster resonance}

\author{Xi-Rong Huang}
\affiliation{Fujian Key Laboratory of Quantum Information and Quantum Optics and
Department of Physics, Fuzhou University, Fuzhou, Fujian 350116, China}

\author{Zong-Xing Ding}
\affiliation{Fujian Key Laboratory of Quantum Information and Quantum Optics and
Department of Physics, Fuzhou University, Fuzhou, Fujian 350116, China}

\author{Chang-Sheng Hu}
\affiliation{Fujian Key Laboratory of Quantum Information and Quantum Optics and
Department of Physics, Fuzhou University, Fuzhou, Fujian 350116, China}

\author{Li-Tuo Shen}
\affiliation{Fujian Key Laboratory of Quantum Information and Quantum Optics and
Department of Physics, Fuzhou University, Fuzhou, Fujian 350116, China}

\author{Weibin Li}

\affiliation{School of Physics and Astronomy, University of Nottingham, Nottingham
NG7 2RD, United Kingdom and Centre for the Mathematics and Theoretical
Physics of Quantum Non-equilibrium Systems, University of Nottingham,
Nottingham NG7 2RD, United Kingdom}

\author{Huaizhi Wu} 

\affiliation{Fujian Key Laboratory of Quantum Information and Quantum Optics and
Department of Physics, Fuzhou University, Fuzhou, Fujian 350116, China}

\affiliation{School of Physics and Astronomy, University of Nottingham, Nottingham
NG7 2RD, United Kingdom and Centre for the Mathematics and Theoretical
Physics of Quantum Non-equilibrium Systems, University of Nottingham,
Nottingham NG7 2RD, United Kingdom}

\author{Shi-Biao Zheng} 
\affiliation{Fujian Key Laboratory of Quantum Information and Quantum Optics and
Department of Physics, Fuzhou University, Fuzhou, Fujian 350116, China}

\begin{abstract}
In this paper, we propose a scheme to implement the two-qubit controlled-Z
gate via the Stark-tuned F\"{o}rster interaction of Rydberg atoms, where
the F\"{o}rster defect is driven by a time-dependent electric field of
a simple sinusoidal function while the matrix elements of the dipole-dipole
interaction are time-independent. It is shown that when the system
is initially in a specific state, it makes a cyclic evolution after
a preset interaction time, returning to the initial state, but picks
up a phase, which can be used for realizing a two-atom controlled-Z
gate. Due to the interference of sequential Landau-Zener transitions,
the population and phase of the state is quasi-deterministic after
the cyclic evolution and therefore the gate fidelity is insensitive
to fluctuations of the interaction time and the dipole-dipole matrix
elements. Feasibility of the scheme realized with Cs atoms is discussed
in detail, which shows that the two-qubit gate via Landau-Zener control
can be realized with the state-of-the-art experimental setup.
\end{abstract}
\maketitle

\section{introduction}

Rydberg atoms trapped in optical potentials provide an attractive
physical architecture for quantum information processing \cite{Saffman_RMP_2010}.
Long-range interactions between distant Rydberg atoms can be switched
on and off on demand with focused lasers \cite{Browaeys_MOP_2016}.
After the pioneering work proposed by Jaksch et al. \cite{Jaksch_PRL_2000},
a number of schemes have been proposed to implement quantum gates
with Rydberg atoms using, e.g. full and partial blockade, as well
as antiblockade \cite{M=0000FCller_PRL_2009,Bhaktavatsala_PRA_2014,Keating_PRA_2015).,Isenhower_PRL_2010,Zhang_PRA_2012,M=0000FCller_PRA_2014,Maller_PRA_2015,Su_PRA_2017,Wu_PRA_2017,Theis_PRA_2016,Goerz_PRA_2014,Petrosyan_PRA_2017,Beterov_PRA_2013,Shao_PRA_2017,M=0000F8ller_PRL_2008,Su_PRA_2016}.
A useful way for controlling the interaction is Stark-tuned F\"{o}rster
resonance \cite{F=0000F6rster_1948}, where two pairs of Rydberg states
that allow for dipole transitions in-between can be shifted into resonance
by dc or microwave electric field \cite{Ryabtsev_PRL_2010,Hofferberth_NC}.
The coherent coupling at F\"{o}rster resonance has been recently demonstrated
in experiment \cite{Ravets_NatPhys_2014,Yakshina_PRA2016} and proposed
for implementing quantum logic gates earlier \cite{Lukin_PRL_2001,Wu_PRA_2010}.
To achieve high fidelity gates, one typically has to control F\"{o}rster
resonances precisely, which means these schemes are sensitive to fluctuations
of interatomic distances and intensity of external fields.

To reduce the effect of parameter fluctuations in dynamical control,
Beterov et al. have recently proposed a scheme for realizing a controlled-Z
(CZ) gate based on a double adiabatic passage across the Stark-tuned
F\"{o}rster resonance, enabling complete population transfer and accumulation
of a deterministic phase for the targeted Rydberg pair state \cite{Beterov_PRA_2016}.
However, to avoid manipulating the distance dependent matrix elements
of the dipole-dipole interaction, the modulation function of the electric
field applied for the Stark-tuning requires to have a power law relation
with respect to the gate operation time, which may increase experimental
complexity nevertheless \cite{Beterov_PRA_2016,Beterov_PRA_2018}. 

Coherent population transfer in a two-level system can be realized
alternatively via periodic sweeping of the interaction-induced avoided
level crossing under the control of an external field, giving rise
to the Landau-Zener (LZ) transitions \cite{Beterov_PRA_2018,Fioretti_PRL_1999,Saquet_PRL_2010,Zhang_PRL_2018,Ditzhuijzen_PRA_2009}
and the Landau\textendash Zener\textendash Stückelberg (LZS) oscillations
\cite{Ditzhuijzen_PRA_2009}. The latter is also referred to as LZS
interference since repeated passages through an avoided crossing act
as an atomic interferometer \cite{Berns_PRL_2006}, causing interference
among different components of the atomic superposition state. If more
than one crossing is involved and the dynamics is overall coherent,
then transition paths can interfere according to the phases accumulated
between subsequent crossings \cite{Basak_PRL_2018}. The LZ and LZS
dynamics have been experimentally demonstrated with Rydberg atoms
\cite{Fioretti_PRL_1999,Saquet_PRL_2010,Zhang_PRL_2018,Ditzhuijzen_PRA_2009,Tretyakov_PRA2014}. 

Inspired by Ref. \cite{Beterov_PRA_2016}, in this paper, we propose
a scheme for implementation of two-qubit logic gates based on the
LZ control of the F\"{o}rster interaction. The dipole-dipole matrix elements
for the coupling between the two Rydberg atoms remain constant during
the gate operation, while the F\"{o}rster defect is periodically modulated
such that the interatomic interaction oscillates between van der Waals
and dipolar shapes. The dynamics of the Rydberg pair states subjected
to the F\"{o}rster interaction is described by the LZS theory, and is
discussed in strong, weak, and intermediate driving regimes, respectively.
Our result shows that a two-qubit CZ gate with high fidelity can be
implemented based on a quasi-deterministic population transfer and
phase accumulation, which can be much less sensitive to the fluctuations
of the gate operation time and the dipole-dipole matrix elements compared
with the scheme based on direct coherent coupling. Furthermore, the
adiabatic passage based scheme (cf. Ref. \cite{Beterov_PRA_2016})
is implemented through sequentially applying two nonlinear driving
pulses, whose intensities and durations need to be exactly identical
and follow power-law dependence on time. Thus, it may be sensitive
to time deviations, see further discussion in Sec. IV. However, the
intensity of the driving field in the LZS based scheme is a simple
harmonic function of time and therefore the experimental complexity
can be greatly reduced. 

\begin{figure}
\centering{}\includegraphics[width=1\columnwidth]{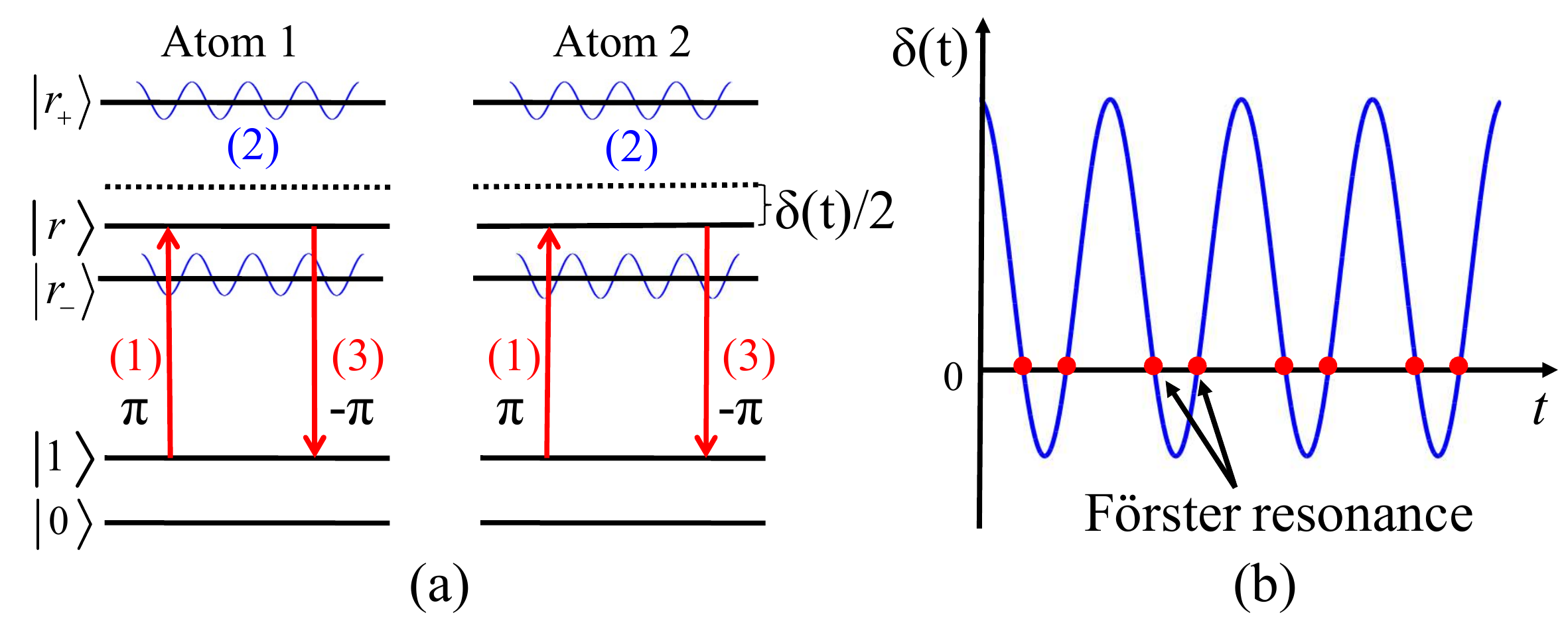}\caption{\label{Fig.a. two-atom scheme}(Color online) (a) Scheme of a CZ gate
based on Landau-Zener dynamics. Two atoms are first excited to the
Rydberg state $|r\rangle$, followed by a harmonic driving that shifts
the neighboring Rydberg levels $|r_{\pm}\rangle$ back and forth modulating
the F\"{o}rster resonance. The atoms are e finally deexcited to the ground
state $|1\rangle$. The phase shift is accumulated if both atoms are
initially prepared in the state $|1\rangle$, and $|0\rangle${\large{}
}is an auxiliary computational state. \label{Fig.b. avoiding crossing }(b)
Time dependence of the energy defect from the F\"{o}rster resonance $\delta\left(t\right)=\delta_{0}+\hbar A\cos\left(\omega t+\phi\right)$,
where the red dots denote the system passing through the F\"{o}rster resonance
induced by periodic modulation.}
\end{figure}

\section{MODEL AND SCHEME}

As shown in Fig. \ref{Fig.a. two-atom scheme}(a), we consider two
identical Rydberg atoms individually trapped in optical tweezers.
Each one has two ground states $|0\rangle$ and $|1\rangle$, which
represent the logic states of the corresponding qubit, and three Rydberg
states $|r\rangle$, $|r_{+}\rangle$ and $|r_{-}\rangle$. The transitions
between the Rydberg levels $|r\rangle$ and $|r_{+}\rangle$ ($|r_{-}\rangle$)
are dipole allowed and the bare energies of the Rydberg pair states
$|r\rangle_{1}|r\rangle_{2}$, $|r_{+}\rangle_{1}|r_{-}\rangle_{2}$,
and $|r_{-}\rangle_{1}|r_{+}\rangle_{2}$ are almost degenerate. The
pair states are coupled by the dipolar interaction based on the F\"{o}rster
process

\begin{equation}
|r\rangle_{1}|r\rangle_{2}\leftrightarrow|r_{+}\rangle_{1}|r_{-}\rangle_{2}+|r_{-}\rangle_{1}|r_{+}\rangle_{2},\label{eq:Forster_channel}
\end{equation}
with the Rabi frequency (strength) $V_{DD}/2$ and the F\"{o}rster defect
$\delta=E_{r_{+}r_{-}}-E_{rr}$, which can be modulated by an external
electric field. Note that the two atoms excited to different Rydberg
states (e.g. $|r_{a}\rangle$ and $|r_{b}\rangle$) may experience
F\"{o}rster resonances as well and the interaction channel will be revised
as $|r_{a}\rangle_{1}|r_{b}\rangle_{2}\leftrightarrow|r_{+}\rangle_{1}|r_{-}\rangle_{2}$
(or $|r_{b}\rangle_{1}|r_{a}\rangle_{2}\leftrightarrow|r_{-}\rangle_{1}|r_{+}\rangle_{2}$
) \cite{Hofferberth_NC,Beterov_PRA_2016}. The F\"{o}rster interactions
can be found in both Rubidium and Cesium Rydberg atoms, see Sec. IV
for further discussion.

The two-qubit controlled-Z gate is implemented through LZ control
of the F\"{o}rster defect in three steps. Step (1): The two atoms are
simultaneously excited to the Rydberg state $|r\rangle$ by a short
$\pi$ pulse when they are in the state $|1\rangle$, and the electric
field is tuned far away from F\"{o}rster resonance {[}see Fig.\ref{Fig.b. avoiding crossing }(b){]}
so that the atomic pair transitions $|r\rangle_{1}|r\rangle_{2}\rightarrow|r_{+}\rangle_{1}|r_{-}\rangle_{2}$
($|r_{-}\rangle_{1}|r_{+}\rangle_{2}$) are adiabatically inhibited.
Step (2): By applying a time-dependent sinusoidal electric field of
radio frequency, the energy defect $\delta\left(t\right)$ is tuned
to zero periodically and the system transits in between the pair Rydberg
states by passing through the avoided level crossing induced by the
resonant dipole-dipole interaction (i.e. the F\"{o}rster resonance). This
results in coherent population transfer of the system states from
$|r\rangle_{1}|r\rangle_{2}$ to $(|r_{+}\rangle_{1}|r_{-}\rangle_{2}+|r_{-}\rangle_{1}|r_{+}\rangle_{2})/\sqrt{2}$
and then back to $|r\rangle_{1}|r\rangle_{2}$, accompanied by accumulation
of a phase shift $\pi$. Note that the coherent population transfer
can be realized as well for the energy defect $\delta\left(t\right)$
being much larger than the inherent dipole-dipole matrix elements,
however, this is non ideal for realization of the CZ gate, see further
discussion below. Step (3): A de-excitation pulse (the second $\pi$
pulse) is applied to the two atoms, transforming the doubly excitation
state back to $|1\rangle_{1}|1\rangle_{2}$. Provided that one of
the atoms is initially in the state $|0\rangle$, no phase shift can
occur because F\"{o}rster resonances are not present. Consequently, the
system evolution is equivalent to the CZ gate:
\begin{gather}
|0\rangle_{1}|0\rangle_{2}\longrightarrow|0\rangle_{1}|0\rangle_{2},\quad|0\rangle_{1}|1\rangle_{2}\longrightarrow|0\rangle_{1}|1\rangle_{2},\nonumber \\
|1\rangle_{1}|0\rangle_{2}\longrightarrow|1\rangle_{1}|0\rangle_{2},\quad|1\rangle_{1}|1\rangle_{2}\longrightarrow-|1\rangle_{1}|1\rangle_{2}.\label{eq:CZ_gate}
\end{gather}

\section{LZS CONTROL OF RYDBERG PAIR STATES}

\begin{figure}
\begin{centering}
\includegraphics[width=0.7\columnwidth]{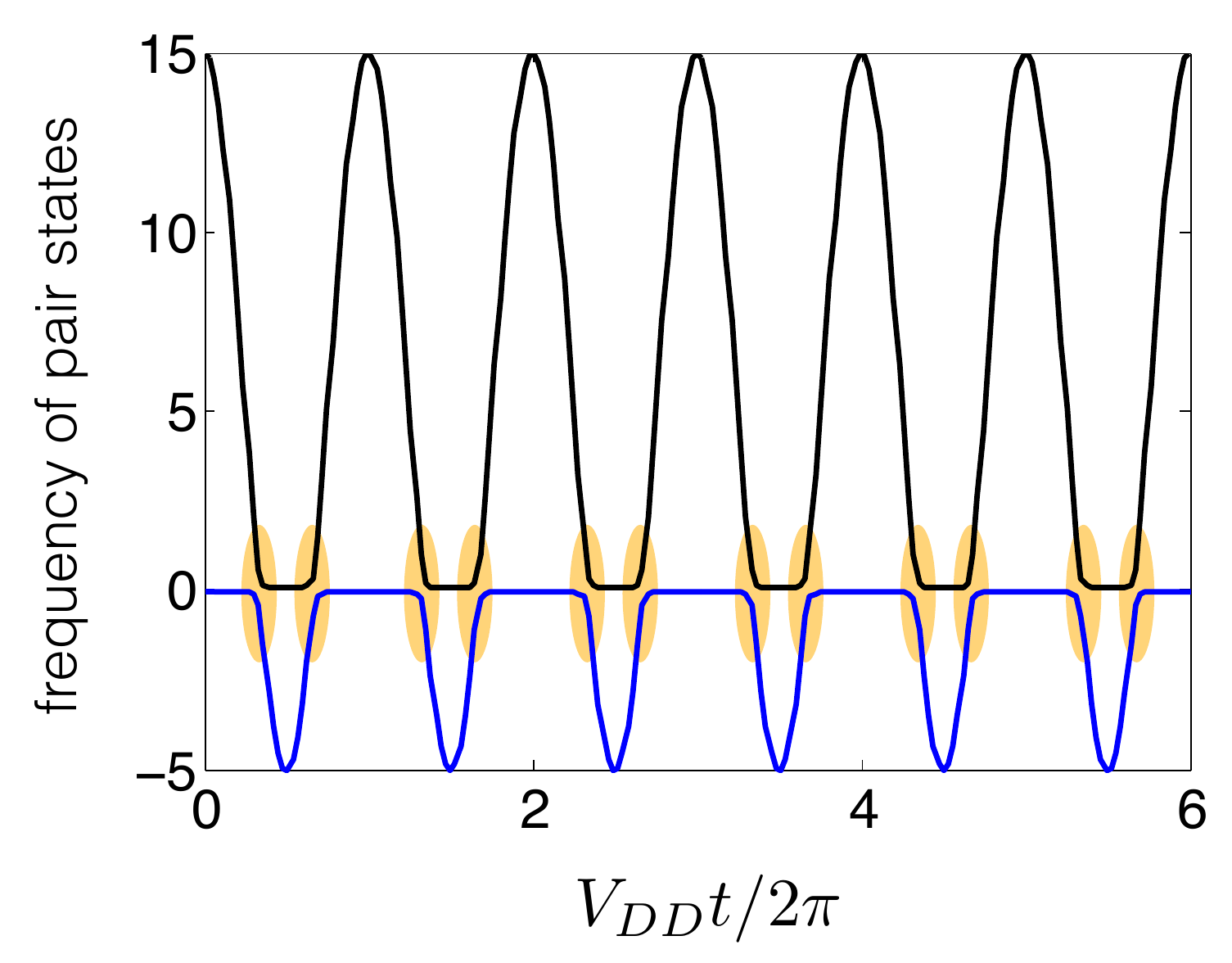}
\par\end{centering}
\caption{\label{fig:Freq_PairState}(Color online) Frequencies of the two collective
states $|\pm\rangle$ as a function of rescaled time. The shaded regions
indicate the avoided crossings resulted from resonant dipole-dipole
interactions. Parameters are $V_{DD}=1$ and ($A$, $\delta_{0}$,
$\omega$)$/V_{DD}$ = (10, 5, 1). }
\end{figure}

To illustrate the realization of the essential transformation $|1\rangle_{1}|1\rangle_{2}\longrightarrow-|1\rangle_{1}|1\rangle_{2}$
(i.e. $|r\rangle_{1}|r\rangle_{2}\longrightarrow-|r\rangle_{1}|r\rangle_{2}$)
more clearly, we reduce the F\"{o}rster process by the coupling between
two symmetric pair states for the two atoms: $|g\rangle\equiv|r\rangle_{1}|r\rangle_{2}$
and $|e\rangle\equiv(|r_{+}\rangle_{1}|r_{-}\rangle_{2}+|r_{-}\rangle_{1}|r_{+}\rangle_{2})/\sqrt{2}$.
The Hamiltonian for the F\"{o}rster defect being driven by a time-varying
radio-frequency (rf) field is then given by ($\hbar=1$)

\begin{equation}
\begin{array}{ccc}
\hat{H}(t) & = & -\frac{1}{2}\left(\begin{array}{cc}
0 & V_{DD}\\
V_{DD} & 2\delta\left(t\right)
\end{array}\right)\end{array}\label{eq:Hamil}
\end{equation}
with 
\[
\delta\left(t\right)=\delta_{0}+A\cos\left(\omega t+\phi\right),
\]

\noindent where the bare energy of the state $|g\rangle$ is set to
zero, and the coupling strength between the two new defined basis
states (i.e. the energy splitting of the avoiding crossing) is assumed
to be independent of time. A key element here is the time varying
detuning $\delta\left(t\right)$, which is a periodic function with
offset $\delta_{0}$, amplitude $A$, and frequency $\omega$. For
simplicity, we take the phase $\phi=0$ in the following. The eigenenergies
of $\hat{H}(t)$ under periodic modulation, which correspond to the
frequencies of the two collective states $|+\rangle=\text{cos}\theta(t)|e\rangle+\text{sin}\theta(t)|g\rangle$
and $|-\rangle=\text{cos}\theta(t)|g\rangle-\text{sin}\theta(t)|e\rangle$
with $\theta(t)=\text{tan}^{-1}(V_{DD}/\delta)/2$, show avoided crossings
while the F\"{o}rster defect is tuned towards resonance, as shown in Fig.
\ref{fig:Freq_PairState}. The effective model (\ref{eq:Hamil}) without
involving pulse shaping of the coupled Rabi frequency is closely related
to Rydberg experiments, where the matrix elements of the dipole-dipole
coupling between two Rydberg atoms are determined by the interatomic
distance and the orientation of the individual dipole, and cannot
be continuously changed in short time scales.

We then rewrite the system Hamiltonian (\ref{eq:Hamil}) by separating
it into time-independent and time-dependent driving parts:

\begin{equation}
\hat{H}(t)=\hat{H}_{0}+\hat{H}_{d}\left(t\right)
\end{equation}
with

\begin{equation}
\hat{H}_{0}=-\delta_{0}|e\rangle\langle e|-\frac{1}{2}V_{DD}(|g\rangle\langle e|+|e\rangle\langle g|),
\end{equation}

\begin{equation}
\hat{H}_{d}\left(t\right)=-A\cos\omega t|e\rangle\langle e|.
\end{equation}

\noindent \begin{flushleft}
In the rotating frame of $\hat{H}_{d}\left(t\right)$, $\hat{H}_{0}$
can be transformed into
\begin{align}
\hat{H^{'}}(t) & =\hat{U}(t)\hat{H}_{0}\hat{U}^{\dagger}(t)-i\hat{U}(t)\dot{\hat{U}}^{\dagger}(t)\nonumber \\
 & =-\frac{1}{2}\left(\begin{array}{cc}
0 & V_{DD}e^{-i(A/\omega)\sin\omega t}\\
V_{DD}e^{i(A/\omega)\sin\omega t} & 2\delta_{0}
\end{array}\right)\label{eq: rotated_H}
\end{align}
\par\end{flushleft}

\noindent with the operator $\hat{U}(t)$ being

\begin{equation}
\hat{U}(t)=\exp\left(-i\int_{0}^{t}\hat{H}_{d}\left(t\right)dt\right)=\exp\left[i\left(\frac{A}{\omega}\sin\omega t\right)|e\rangle\langle e|\right],
\end{equation}

\noindent which also maps the wave function in the reference frame
$|\psi\rangle$ onto $|\psi^{'}\rangle$ via $|\psi^{'}\rangle=\hat{U}(t)|\psi\rangle$,
following the Schrödinger equation $i\frac{d}{dt}|\psi^{'}\rangle=\hat{H^{'}}(t)|\psi^{'}\rangle$
. The Hamiltonian (\ref{eq: rotated_H}) after making use of the Jacobi-Anger
expansion

\[
e^{ix\sin\tau}=\stackrel[n=-\infty]{\infty}{\sum}J_{n}(x)e^{in\tau}
\]

\noindent takes the form

\begin{equation}
\hat{H}^{\prime}(t)=-\frac{1}{2}\left(\begin{array}{cc}
0 & \stackrel[n=-\infty]{\infty}{\sum}\Omega_{n}e^{-in\omega t}\\
\stackrel[n=-\infty]{\infty}{\sum}\Omega_{n}^{*}e^{in\omega t} & 2\delta_{0}
\end{array}\right),\label{eq:H_eff}
\end{equation}
where the periodic energy defect has effectively modified
the Rabi coupling $\Omega_{n}=V_{DD}J_{n}(\frac{A}{\omega})$ with
$J_{n}(\frac{A}{\omega})$ being the $n$th order Bessel function
of the first kind. In addition, the resonance condition $\delta_{0}=m\omega$,
which describes $|m|$-rf-photon transition process between the two
collective states assisted by the rf driving field, can be identified
by examining the time independent term of the non-diagonal elements
if the rotating wave approximation is made \cite{Ashhab_PRA_2007}. 

The LZ dynamics determined by $\hat{H}^{\prime}$ strongly depends
on the driving parameters $A$, $\delta_{0}$, and $\omega$, which
must satisfy two conditions to implement a robust CZ gate. First,
the system should be able to make a full cycle of Rabi-like oscillation
between the two collective states and accumulate a $\pi$ phase, which
requires the F\"{o}rster defect $\sim\delta_{0}$ to be an integer multiple
of the rf-photon frequency for $\delta_{0}\gg V_{DD}$, and the frequency
of the oscillation $\sim V_{DD}J_{n}(A/\omega)$ should be as large
as possible such that the gate operation is decoherence-resistant
\cite{Ashhab_Review_RP2010}. Second, the oscillatory period needs
to be an integer multiple of the time period $\tau_{d}=2\pi/\omega$
of a complete LZ passage, which makes the population and phase of
the state $|g\rangle$ after the cyclic evolution robust to imperfect
timing for a slow passage ($\omega/V_{DD}\sim1$). In the following,
we study three different regimes of the driving parameters and focus
on the situations that the F\"{o}rster defect between the pair of Rydberg
states is large compared with the dipole-dipole matrix elements (i.e.
$\delta_{0}\gg V_{DD}$).

\begin{figure}
\centering{}\includegraphics[width=0.7\columnwidth]{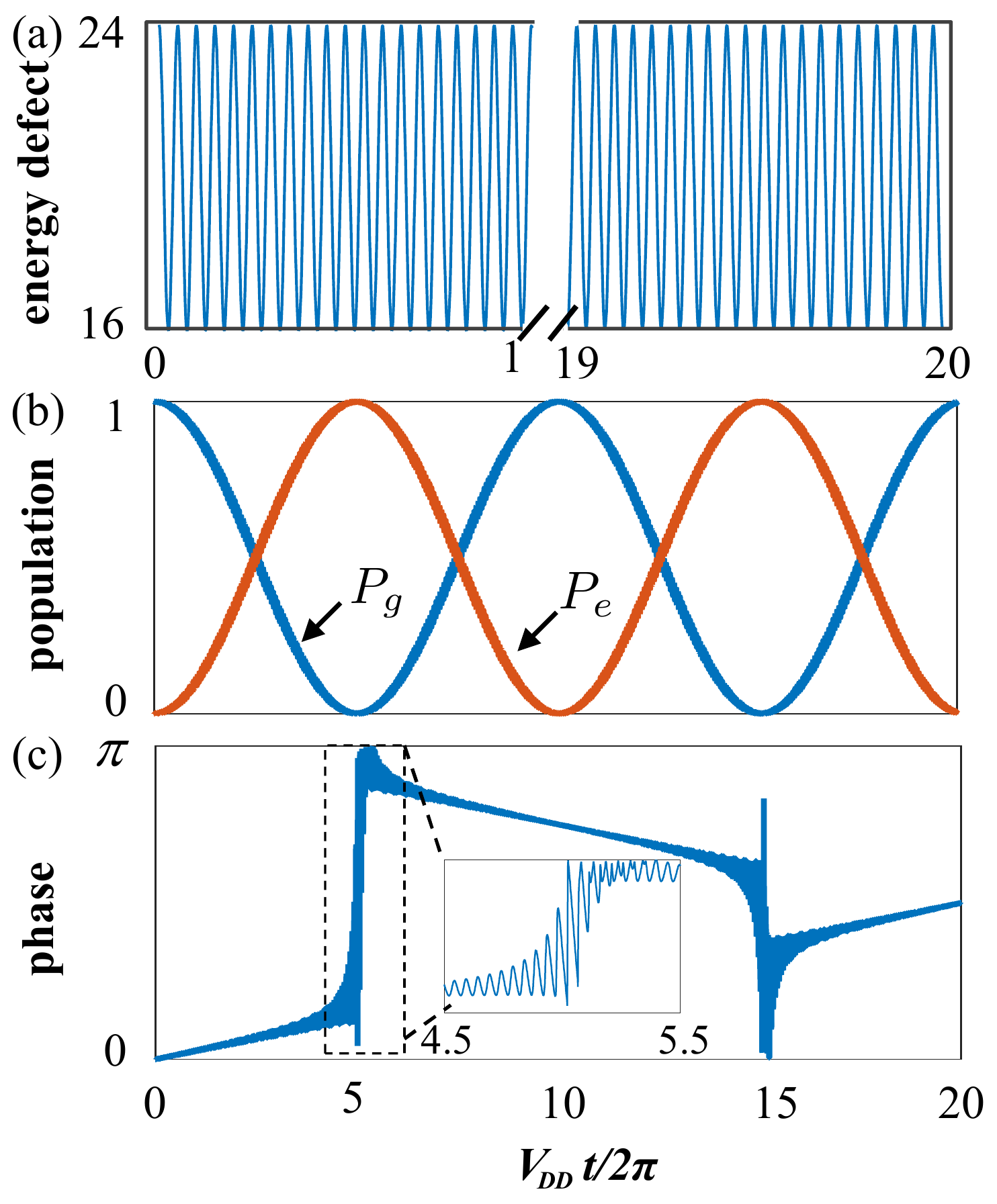}\caption{\label{fig:weak_drive}(Color online) (a) Time-dependent energy defect
$\delta\left(t\right)$ as a function of dimensionless rescaled time
$V_{DD}t$. (b) Evolutions of the populations of the state $|g\rangle$
($P_g$) and the excited state $|e\rangle$ ($P_e$) for the
system initially in the state $|g\rangle$. (c) Time dependent phase
of the state{\large{} }$|g\rangle${\large{}.} We fix units of $V_{DD}=1$
and set ($A$, $\delta_{0}$, $\omega$)$/V_{DD}$= (4, 20, 20). }
\end{figure}

\begin{figure*}[t]
\begin{centering}
\includegraphics[width=1.0\textwidth]{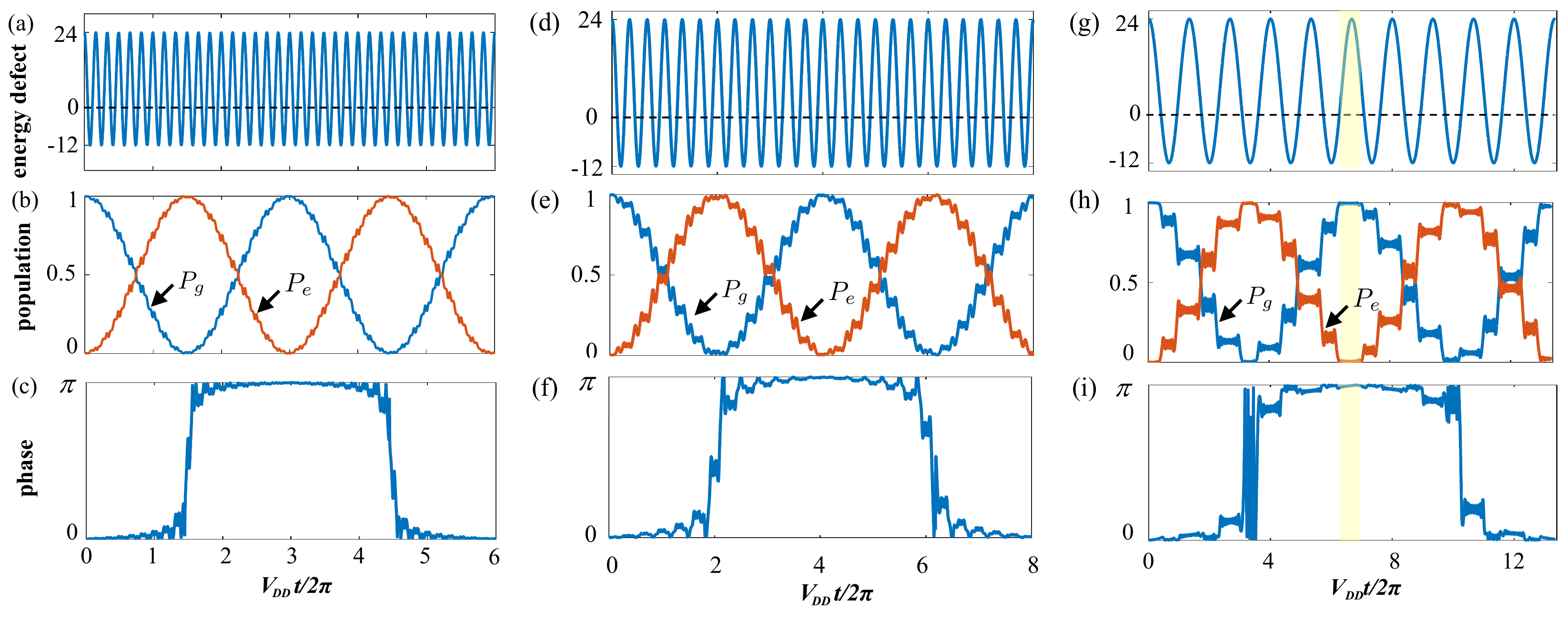}
\par\end{centering}
\caption{\label{fig: strong_drive_fast}(Color online) (a), (d), (g) Time-dependent
energy defect $\delta\left(t\right)$ as a function of dimensionless
rescaled time $V_{DD}t$. (b), (e), (h) Evolutions of the populations
of the state $|g\rangle$ ($P_g$) and the excited state $|e\rangle$ ($P_e$) for the system initially in the state $|g\rangle$. (c), (f),
(i) Time dependent phase of the state{\large{} }$|g\rangle${\large{}.}
We fix units of $V_{DD}=1$, and set $\left(A,\delta_{0},\omega\right)/V_{DD}=\left(18,6,6\right)$
in (a)-(c), $\left(A,\delta_{0},\omega\right)/V_{DD}=\left(18,6,3\right)$
in (d)-(f), $\left(A,\delta_{0},\omega\right)/V_{DD}=\left(18,6,0.75\right)$
in (g)-(i), respectively.}
\end{figure*}

\textit{Weak driving}. The weak driving regime refers to $A\ll E_{q}\equiv\sqrt{\delta_{0}^{2}+V_{DD}^{2}}$,
under which, the single-rf-photon resonant transition from $|g\rangle$
to $|e\rangle$ occurs for $\omega=\sqrt{\delta_{0}^{2}+V_{DD}^{2}}$
and the frequency of the Rabi oscillations is given by $\Omega_{eg}\equiv A\sin[\tan^{-1}(V_{DD}/\delta_{0})]/2$.
If the system is initially in the state $|g\rangle$ with a large
static defect $\delta_{0}\gg V_{DD}$, the valid approximation $\sin[\tan^{-1}(V_{DD}/\delta_{0})]\approx V_{DD}/\delta_{0}$
can be made to the Rabi frequency, giving rise to $\Omega_{eg}\approx AV_{DD}/2\delta_{0}$.
As a special case, this also describes the Autler-Townes splitting
at small driving amplitude $A$, and can be analytically calculated
by simply truncating the series of $\Omega_{n}$ up to $n=1$ (corresponding
to the rotating wave approximation), which leads to \cite{Marquardt_LZS}
\begin{eqnarray}
\hat{H}^{'}(t) & \approx & -\frac{1}{2}\left(\begin{array}{cc}
0 & V_{DD}J_{0}(\frac{A}{\omega})\\
V_{DD}J_{0}(\frac{A}{\omega}) & 2\delta_{0}
\end{array}\right)\nonumber \\
 &  & -\frac{1}{2}\left(\begin{array}{cc}
0 & V_{DD}J_{1}(\frac{A}{\omega})e^{i\omega t}\\
V_{DD}J_{1}(\frac{A}{\omega})e^{-i\omega t} & 0
\end{array}\right).\label{eq: H_eff_1st}
\end{eqnarray}
Using the interaction picture representation, we can then find the
effective transition frequency $\sqrt{\delta_{0}^{2}+V_{DD}^{2}J_{0}^{2}(\frac{A}{\omega})}$
and the Rabi frequency $V_{DD}J_{1}(\frac{A}{\omega})\approx AV_{DD}/2\delta_{0}$
of the two-level system by using the approximation $J_{n}(\frac{A}{\omega})\sim\frac{(A/\omega)^{n}}{2^{n}n!}$
for $A/\omega\ll1$. According to Eq. (\ref{eq: H_eff_1st}), if the
system is initially in the state $|g\rangle$, it will undergo Rabi
oscillations between $|g\rangle$ and $|e\rangle$ and return to $|g\rangle$
after a full Rabi cycle, but pick up a phase $\varphi$. This is illustrated
in Fig. \ref{fig:weak_drive}, but which shows $\varphi$ is generally
not equal to $\pi$, so that a two-qubit controlled-Z gate cannot
be realized in this regime. 

\textit{Strong driving.} We next turn to the case of strong driving
with $(A-\delta_{0})\gg V_{DD}$, where the system repeatedly traverses
the F\"{o}rster resonance and hardly spends any time in the degeneracy
point \cite{Ashhab_PRA_2007}. To gain the insight, we perform a further
transformation $exp(i\delta_{0}|e\rangle\langle e|t)$, transforming
the Hamiltonian of Eq. (\ref{eq:H_eff}) to
\begin{equation}
\hat{H}_{I}(t)=-\frac{1}{2}\stackrel[n=-\infty]{\infty}{\sum}\left(\begin{array}{cc}
0 & \Omega_{n}e^{i(\delta_{0}-n\omega)t}\\
\Omega_{n}^{*}e^{-i(\delta_{0}-n\omega)t} & 0
\end{array}\right).\label{eq: H2_eff}
\end{equation}
Under the condition $\delta_{0}=m\omega$, the driving associated
with the effective frequency component with $n=m$ is in resonance,
corresponding to a $|m|$-rf-photon process. In the high frequency
limit, where the frequency of the external driving is much larger
than the effective Rabi frequencies associated with the other frequency
components, i.e., $\omega\gg\Omega_{n}$ ($n\neq m$), all time-dependent
fast oscillating terms ($\sim e^{i(m-n)\omega t}$) can be neglected.
As a consequence, the system dynamics is reduced to the resonant driving
of a two-level system with the Rabi frequency $\Omega_{n}=V_{DD}|J_{n}(\frac{A}{\omega})|$,
as shown in Figs. \ref{fig: strong_drive_fast} (a)-\ref{fig: strong_drive_fast}(c).
In this case, the two-level system can make a cyclic evolution, and
return to the initial state $|g\rangle$, picking up a phase of $\pi$.
But this occurs almost at a specific moment $T_{cz}\sim1/V_{DD}|J_{1}(\frac{A}{\omega})|$
since the time interval between subsequent transition events is of
the order of half driving period $1/2\omega$, which is short here.
Thus, the evolutional dynamics in analogous to the coherent resonant
coupling scheme requires precise control of the rescaled time. In
addition, we note that there is a special situation with the parameters
$\delta_{0}=0$ and $n=0$, in which the system can transit between
$|e\rangle$ and $|g\rangle$ with full conversion via the LZ control,
however, here we focus on the general case of a finite F\"{o}rster defect. 

\begin{figure*}[t]
\centering{}\includegraphics[width=1.0\textwidth]{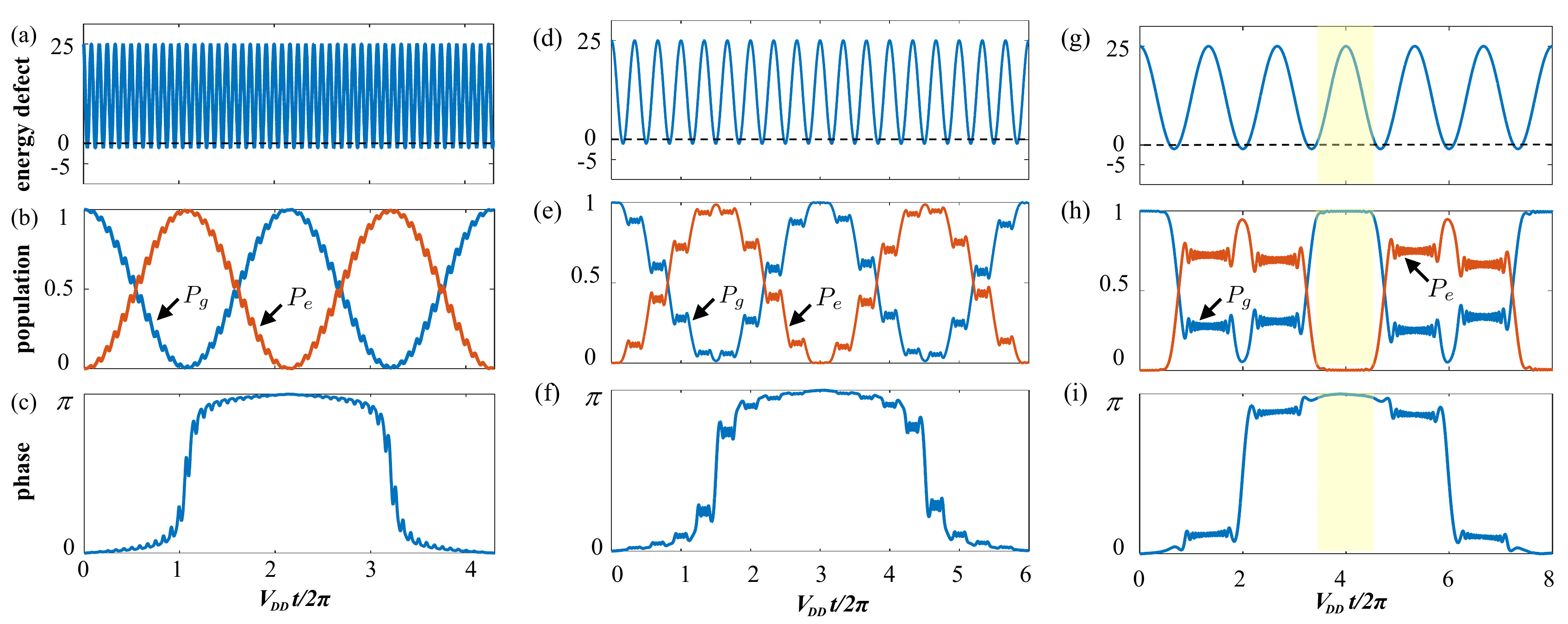}\caption{\label{fig_Intermed_drive}(Color online) (a), (d), (g) Time-dependent
energy defect $\delta\left(t\right)$ as a function of dimensionless
rescaled time $V_{DD}t$. (b), (e), (h) Evolutions of the populations
of the state $|g\rangle$ ($P_g$) and the excited state $|e\rangle$ ($P_e$) for the system initially in the state $|g\rangle$. (c), (f),
(i) Time dependent phase of the state $|g\rangle$.
We fix units of $V_{DD}=1$, and set $\left(A,\delta_{0},\omega\right)/V_{DD}=\left(13,12,12\right)$
in (a)-(c), $\left(A,\delta_{0},\omega\right)/V_{DD}=\left(13,12,3\right)$
in (d)-(f), $\left(A,\delta_{0},\omega\right)/V_{DD}=\left(13,12,0.75\right)$
in (g)-(i), respectively.}
\end{figure*}

In the low-frequency situation $\omega\sim\Omega_{m}$, a stepwise
increase or decrease of the population can be found for each time
the system passing through the LZ avoid crossing and the population
has weak oscillations during its stay at each stair. In general, the
system exhibits non-sinusoidal oscillations and can approximately
return to the initial state $|g\rangle$ with the quasi-deterministic
population and phase $\pi$ after a time period $T$, which are exactly
multiple of the driving period, e.g. $V_{dd}T\sim2\pi\times4$ with
$T/\tau_{d}=12$ in Figs. \ref{fig: strong_drive_fast}(d)-\ref{fig: strong_drive_fast}(f)
and $V_{dd}T\sim2\pi\times7$ with $T/\tau_{d}=5$ in Figs. \ref{fig: strong_drive_fast}(g)-\ref{fig: strong_drive_fast}(i).
The duration for the system staying in $|g\rangle$ after an oscillation
period is again determined by the time interval for two sequential
sweeping of the avoided crossing, as indicated in Figs. \ref{fig: strong_drive_fast}(g)-\ref{fig: strong_drive_fast}(i).
Mathematically, this is due to the fact that the single resonant transition
with $\delta_{0}=m\omega$ is not a good approximation any more in
the low frequency regime, where the ``noise channels'' contribute
to the Rabi coupling between the two basis states if the corresponding
Rabi frequencies of the non-resonant components are comparable to
the detunings, i.e. $(n-m)\omega\sim\Omega_{n}$. This feature makes
the gate dynamics robust against certain amount of time deviation
and parameter fluctuation. 

\textit{Intermediate driving.} Finally, we look into the regime with
$A\simeq\delta_{0}$, where the system reaches the F\"{o}rster resonance
around the turning point of the harmonic driving, but stays for a
longer time at the avoided crossing compared with the case of strong
driving. To illustrate the performance of the controlled-Z gates in
this regime, we perform numerical simulations with three sets of parameters,
as shown in Fig. \ref{fig_Intermed_drive}. The results show the system
exhibits the LZ transition behavior similar to that under the strong
driving for both high frequency and slow passage limits. The distinct
features of the dynamics of the system in this regime are that it
can return to the initial state $|g\rangle$ and pick up a $\pi$-phase
after a relative short time, and that the evolution can be frozen
for a relative long duration {[}see the shaded area in Figs. \ref{fig_Intermed_drive}(g)-\ref{fig_Intermed_drive}(i){]}.
These features enable implementation of a high-fidelity controlled-Z
gate that is robust to parameter fluctuations. 

The reason why the performance of the CZ gate in the intermediate
driving regime is better than that under strong driving can be explained
as follows. On one hand, the avoided crossing is passed at a slower
speed and the population exchange for each LZ passage is greatly enhanced.
Therefore, the time period for the cyclic evolution and the corresponding
gate operation time is shortened. The effective Rabi frequency for
the cyclic evolution can be estimated by $\Omega_{eff}\sim V_{DD}J_{m}(A/\omega)$,
which under the condition of $m=\delta_{0}/\omega\gg1$ has a maximum
around $A/\delta_{0}\sim1$ \cite{Ashhab_PRA_2007}. On the other
hand, the dynamics is robust to imperfect timing only in the low-frequency
driving limit $\omega/V_{DD}\sim1$ and is however bounded by the
requirement of an integer number of $T/\tau_{d}\sim\omega/\Omega_{eff}$.
Thus, in the intermediate driving regime {[}see Figs. \ref{fig_Intermed_drive}(g)-\ref{fig_Intermed_drive}(i){]},
the time evolution of the population and phase can be frozen for almost
a complete driving period $2\pi/\omega$ under the minimal driving
frequency $\omega/\Omega_{eff}\sim3$. As a consequence, the system
does not oscillate back and forth between the two collective states
for a smaller driving frequency.

\begin{figure}
\centering{}\includegraphics[width=0.8\columnwidth]{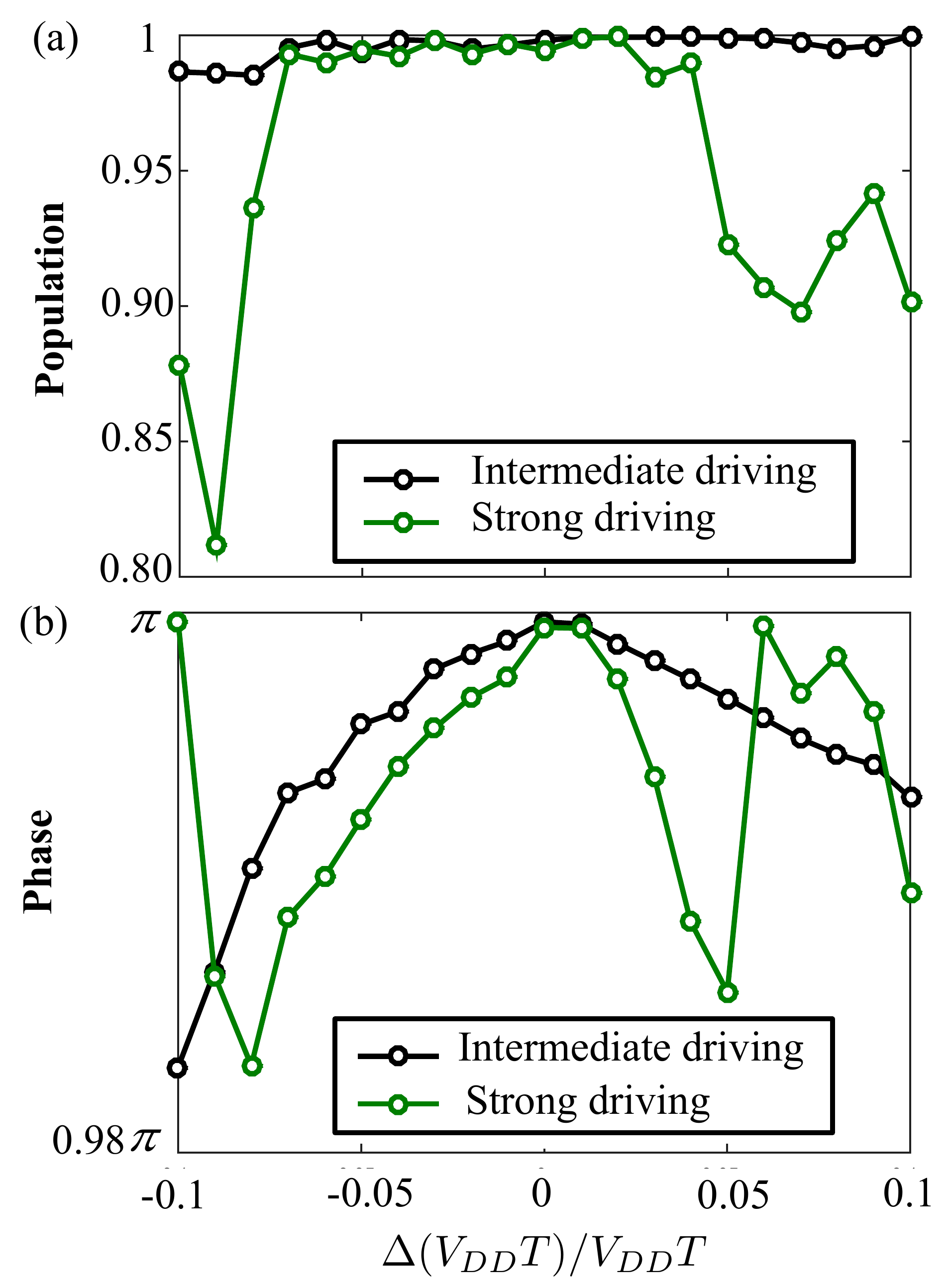}\caption{\label{fig_comparison}(Color online) (a) Population of the state
$|g\rangle$ for the system initially in the state $|g\rangle$ against
the dimensionless rescaled time deviation from the preset interaction
time in the intermediate (black line) and strong (green line) driving
regimes. (b) Phase of $|g\rangle$ against the dimensionless rescaled
time deviation. Parameters are the same as those in Fig. \ref{fig: strong_drive_fast}(g)-(i)
(for strong driving) and Fig. \ref{fig_Intermed_drive}(g)-(i) (for
intermediate driving).}
\end{figure}

To compare the stabilities of the gates realized in the strong and
intermediate driving regimes, we plot the corresponding populations
of the state $|g\rangle$ and the acquired phases against the time
deviation in Fig. \ref{fig_comparison}. The results show that both
the population fluctuation and phase fluctuation in the intermediate
driving regime are smaller than those in the strong driving regime.
In the case of intermediate driving with $\delta(V_{DD}T)/(V_{DD}T)\thicksim10\%$,
we find that the population fluctuation is less than 2\%. On the other
hand, the phase acquired in the strong driving regime oscillates between
0.98$\pi$ and $\pi$ within this time deviation, but the phase error
remains less than 0.02$\pi$.

\section{IMPLEMENTATION OF the CZ GATE AND EXPERIMENTAL FEASIBILITY}

\begin{figure}
\centering{}\includegraphics[width=0.8\columnwidth]{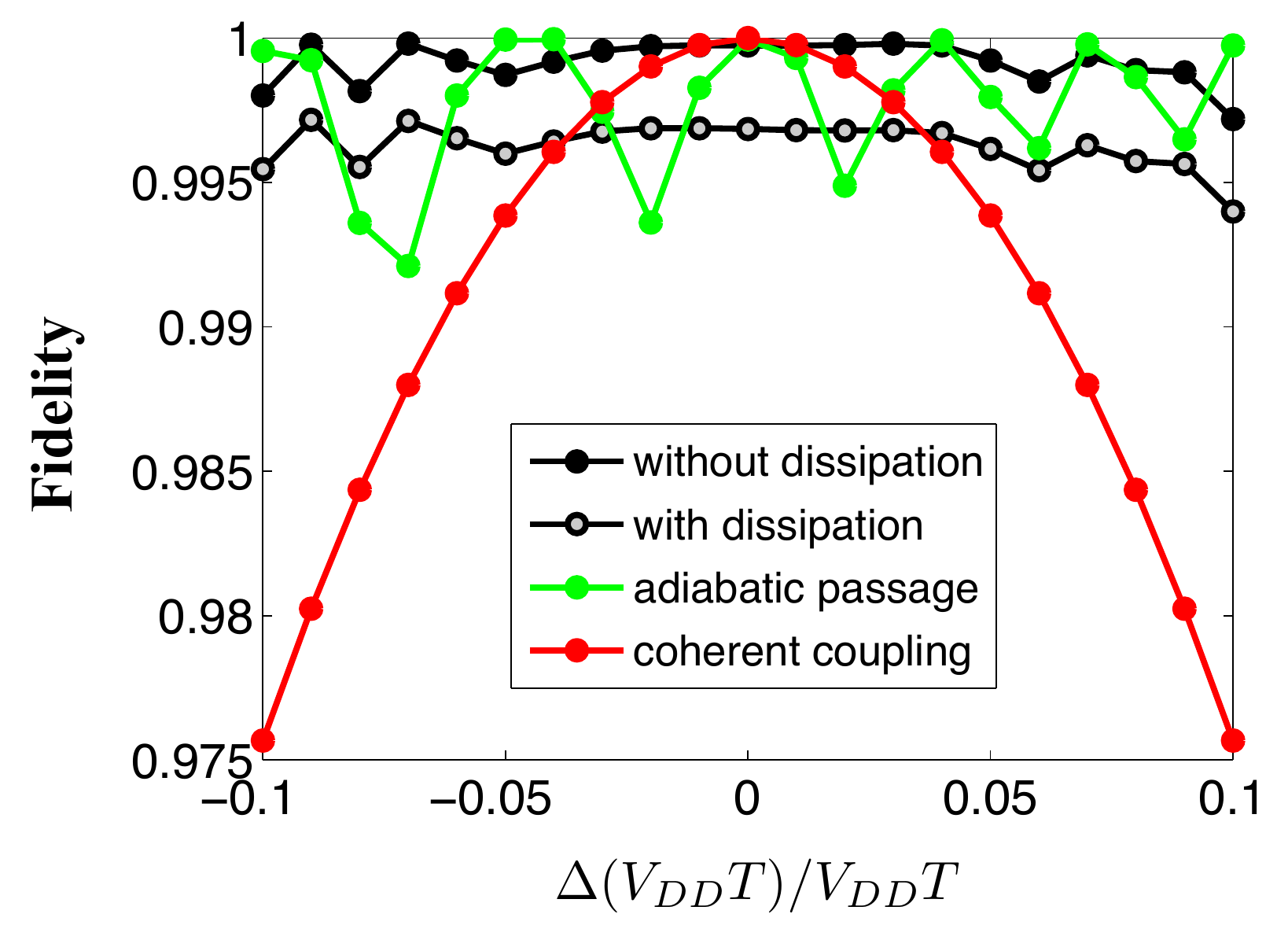}\caption{\label{fig_Med_vs_Coh} (Color online) Fidelities of the CZ gates
versus the dimensionless rescaled time deviation for the schemes through
the LZS approach with and without atomic spontaneous emission, the
direct coherent coupling, and the adiabatic passage in Ref. \cite{Beterov_PRA_2016}.
See the main text for detail. }
\end{figure}

Now we focus on the intermediate driving regime, which allows for
optimal control of the population transfer and the phase accumulation.
To evaluate the performance of the controlled-Z gate (\ref{eq:CZ_gate}),
we take an example, where the input state is $|\psi_{0}\rangle=\frac{1}{2}(|00\rangle+|01\rangle+|10\rangle+|11\rangle)$.
The quality of the output state $|\psi_{f}\rangle$ under nonideal
conditions is characterized by the fidelity, defined as $F=|\langle\psi_{f}|U_{cz}|\psi_{0}\rangle|^{2}$
where $U_{cz}$ is a diagonal matrix with $diag(U_{cz})=(1,1,1,e^{i\pi})$.
We first consider the effect of parameter fluctuation and neglect
the atomic spontaneous emission. Assume that both the first and third
step of the controlled-Z gate operation (i.e. the excitation and de-excitation
of the atoms between $|1\rangle$ and $|r\rangle$) are correctly
implemented, then the fidelity of our scheme is almost perfect ($F=0.9998$)
for $V_{dd}T=2\pi\times4$ with the same parameters as in Fig. \ref{fig_Intermed_drive}.
To examine the advantages of our scheme compared to the approach via
direct coherent coupling at the F\"{o}rster resonance \cite{Ravets_NatPhys_2014},
we then check the robustness of the two schemes to the fluctuation
of the rescaled time for gate operation, as shown in figure \ref{fig_Med_vs_Coh}.
It is clearly verified that the fidelity of our scheme is less sensitive
to the uncertainty of the operation time in contrast to the coherent
coupling method with a similar gate duration. A fidelity as high as
0.995 can be well maintained for a time deviation of $\Delta(V_{DD}T)/(V_{DD}T)\sim10\%$
via the periodic Landau-Zener control, while the result obtained by
the coherent coupling method is about 0.975, which may be affected
by additional errors for tuning the system exactly to the F\"{o}rster
resonance. Furthermore, the nonlinear driving scheme requires precise
control of the two symmetric adiabatic sequences according to the
power-law function $\delta_{1,2}(t)=s_{1}(t-t_{1,2})+s_{2}(t-t_{1,2})^{5}$,
where the optimized set of parameters are $s_{1}/2\pi=-10$ MHz/$\mu\text{s}$,
$s_{2}/2\pi=-2600$ MHz/$\mu\text{s}^{5}$, $T=1.8\text{ }\mu\text{s}$,
$t_{1}=T/4$, $t_{2}=3T/4$ and $V_{DD}/2\pi=2.1$ MHz (cf. Ref. \cite{Beterov_PRA_2016}).
When the operation time for each of the two sequences deviates from
the expected value by $\Delta T/T\sim2\%$, the fidelity will reduce
to $\sim0.995$. In contrast, the LZS based scheme is more robust
to imperfect timing.

In the context of Rydberg experiments, we simply take the example
of the pair-state interaction channel $|90S_{1/2}\rangle+|96S_{1/2}\rangle\rightarrow|90P_{1/2}\rangle+|95P_{1/2}\rangle$
in Cs Rydberg atoms, as previously found by Beterov et al. \cite{Beterov_PRA_2016}.
In this case, the atoms can be addressed individually since they are
excited to different Rydberg states. The F\"{o}rster interaction between
the pairs states has the energy defect $\delta_{0}/2\pi=75.6$ MHz
and the exact F\"{o}rster resonance occurs with the electric field being
tuned to $E=29.75$ mV/cm. On the other hand, to meet the requirement
of the intermediate driving regime (see Fig. \ref{fig_Intermed_drive}),
the dipole-dipole matrix elements for the two atoms along the $z$
axis should be $V_{DD}/2\pi\propto C_{3}/R^{3}\sim3.2$ MHz, which
for this channel is equivalent to the interatomic distance $R=20$
$\mu$m for $C_{3}=-154968$ MHz/$\mu m^{3}$. Correspondingly, the
frequency of the rf driving field is $\omega/2\pi\sim2.4$ MHz, which
is easy to access in experiments. Note that other transition channels
related to this F\"{o}rster resonance are safely neglected because of
the large energy defects, which are the order of several hundred MHz
\cite{Beterov_PRA_2009}. 

The Rydberg states we considered have lifetimes around \cite{Beterov_PRA_2009}
$\tau_{90S}=270\mu s$, $\tau_{96S}=314\mu s$, $\tau_{90P}=361\mu s$,
and $\tau_{95P}=406\mu s$ in the room temperature ($\sim300$ K).
Thus, the two atoms excited to Rydberg states are subjected to atomic
spontaneous emission during the LZS control. The effect of the dissipation
during the gate operation can be evaluated by using the conditional
Hamiltonian 
\begin{equation}
\hat{H}_{cond}=\hat{H}(t)-\frac{i}{2}\underset{r}{\sum}\gamma_{r}(\hat{\sigma}_{rr}^{(1)}+\hat{\sigma}_{rr}^{(2)}),
\end{equation}
where $\hat{\sigma}_{rr}^{(j)}=|r\rangle_{jj}\langle r|$ ($j=1,2$)
and the sum is taken over all Rydberg states of the F\"{o}rster interaction
channel. The numerical estimate with conditional Hamiltonian simply
discards the state components with each of these two atoms going back
to the computational space due to the spontaneous emission, which
may have some overlap with the desired output state. Therefore, it
provides a conservative result on the gate fidelity. As shown in figure
\ref{fig_Med_vs_Coh}, we can see that the spontaneous decay slightly
reduces the gate fidelity, which however, still surpasses 0.995 in
general within 10\% of the deviation of the rescaled time.

\begin{figure}
\includegraphics[width=1\columnwidth]{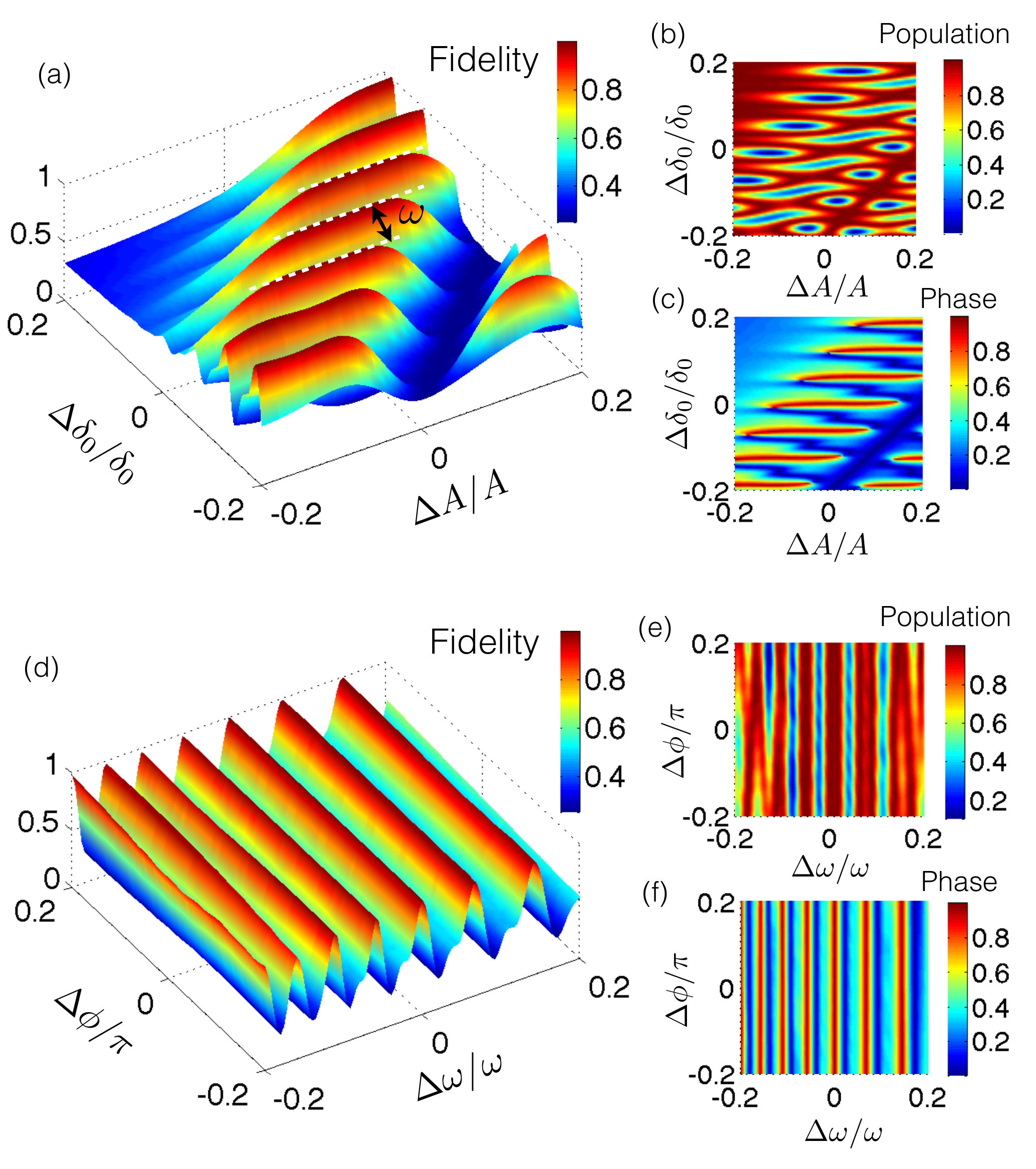}

\caption{\label{fig:Robust_DrivePara}(Color online) Robustness of the fidelity
of the CZ gate, population revival, and phase accumulation (divided
by $\pi$) with respect to deviations of the driving amplitude $\Delta A/A$
and the detuning $\Delta\delta_{0}/\delta_{0}$ {[}(a)-(c){]}, and
deviations of the driving frequency $\Delta\omega/\omega$ and the
initial driving phase $\Delta\phi/\pi$ {[}(d)-(f){]}. Parameters
are the $C_{3}$ coefficient of the dipole-dipole matrix elements
$-154968$ MHz/$\mu m^{3}$, the interatomic distance $R=20$ $\mu\text{m}$,
$(A,\delta_{0},\omega)/2\pi=(83.2,76.8,3.15)$ MHz, and the spontaneous
decay rates $(\gamma_{90S},\gamma_{96S},\gamma_{90P},\gamma_{95P})=(1/270,1/314,1/361,1/406)$
MHz.}
\end{figure}

In terms of the typical parameters with respect to Cs Rydberg atoms,
we now discuss the sensitivity of our scheme to fluctuations of the
driving parameters. As shown in Fig. \ref{fig:Robust_DrivePara}(a),
there are ridges of high fidelity, which have the separation in detuning
exactly given by the driving frequency $\omega$ and correspond to
the multi-rf-photon resonance condition $\delta_{0}=m\omega$. Therefore,
the scheme requires accurate control of the driving frequency (typically
limited by $\Delta\omega\sim\Omega_{eff}/|m|$ \cite{Ashhab_PRA_2007})
although the energy defect $\delta_{0}\pm\Delta\delta_{0}$ allows
the two collective states to transit in between via different resonant
channels {[}see Fig. \ref{fig:Robust_DrivePara}(d){]}. However, both
the population and phase of the state show robustness against small
fluctuations in the driving amplitude {[}see \ref{fig:Robust_DrivePara}(b),
(c){]}, e.g. a deviation of the amplitude $\Delta A/A\approx5\%$
leads to the reduced fidelity $F\approx0.992$, which is comparable
to that of the nonlinear driving scheme with $5\%$ deviations in
$s_{1,2}$. In addition, the population and phase of the state are
highly robust to the initial phase deviation of the rf field {[}see
\ref{fig:Robust_DrivePara}(e), (f){]}.

Furthermore, our theoretical model can be related to the previously
experimental demonstrations of the radio-frequency-assisted F\"{o}rster
resonances $nP_{3/2}+nP_{3/2}\rightarrow nS_{1/2}+(n+1)S_{1/2}$ for
$n<39$ in Rb atoms \cite{Yakshina_PRA2016,Tretyakov_PRA2014}, where
the periodic-driving induced single- and multi-rf-photon transition
can be alternatively explained in terms of the Floquet sidebands \cite{Ditzhuijzen_PRA_2009,Ashhab_Review_RP2010}.
Considering the case that the non-zero F\"{o}rster defect $\delta_{0}$
is now Stark-tuned by the composite electric field consisting of dc
and rf, $E=E_{\text{dc}}+E_{\text{rf}}\text{cos}(\omega t)$. Then,
the time-varying detuning between the pair of collective Rydberg levels
is approximately given by $\delta(t)\approx\delta_{0}^{\prime}+A^{'}\text{cos}(\omega t)$
for weak rf fields $E_{\text{rf}}\ll E_{\text{dc}}$, where $\delta_{0}^{\prime}=\delta_{0}+(\alpha_{nP}-\frac{1}{2}\alpha_{ns}-\frac{1}{2}\alpha_{(n+1)s})E_{\text{dc}}^{2}$,
$A^{'}=2E_{\text{rf}}E_{\text{dc}}(\alpha_{nP}-\frac{1}{2}\alpha_{ns}-\frac{1}{2}\alpha_{(n+1)s})$,
and $\alpha_{nl}$ are the quadratic polarizabilities. By assuming
$\delta_{0}^{\prime}=m\omega$ with $m$ being a non-zero integer,
we find that the robust CZ gate can be implemented in the intermediate
driving regime (i.e. $A^{'}/\delta_{0}^{\prime}\sim1$) iff $\delta_{0}/m\omega\gg3/2$.
For $n=37$ \cite{Yakshina_PRA2016,Tretyakov_PRA2014}, the parameter
regime in Fig. \ref{fig_Intermed_drive}(g)-\ref{fig_Intermed_drive}(i)
corresponds to $\omega/2\pi=1$ MHz, $E_{\text{dc}}\approx1.69$V/cm
with $E_{\text{rf}}/E_{\text{dc}}=0.1$.

\section{CONCLUSION}

In summary, we have proposed an experimental feasible scheme for implementation
of a two-qubit logic gate by modulating the F\"{o}rster resonance with
a periodic driving field. The Stark-tuned F\"{o}rster interaction between
the two pairs of Rydberg states can be regarded as a periodically
driven two-level system, with a time-invariant coupling strength and
a sinusoidal time-dependent detuning. The results show that the gate
can be accomplished within an operation time comparable with that
required by the method based on double adiabatic passages \cite{Beterov_PRA_2016},
and in contrast to the coherent coupling scheme, the gate fidelity
is much less sensitive to the fluctuations of the interaction time
and the motion-sensitive dipole-dipole matrix elements due to the
sequential Landau-Zener transitions. We numerically analyze the implementation
of this gate with the realistic F\"{o}rster interaction channel in
Cs Rydberg atoms, and the results demonstrate its
performance is insensitive to both the time fluctuations and atomic
spontaneous emission, confirming its promise in quantum information
processing.
\begin{acknowledgments}
W.L. acknowledges support from the UKIERI-UGC Thematic Partnership
No. IND/CONT/G/16-17/73, and EPSRC Grant No. EP/M014266/1 and EP/R04340X/1.
L.T.S., H.W. and S.B.Z are supported by the National Natural Science
Foundation of China under Grants No. 11774058, No. 11674060, No. 11874114 and No.
11705030, the Natural Science Foundation of Fujian Province under
Grant No. 2017J01401, and the Qishan fellowship of Fuzhou University. H.W. acknowledges particularly the financial
support by the China Scholarship Council for the academic visit of
the University of Nottingham. 
\end{acknowledgments}

\end{document}